\documentclass{emulateapj}

\usepackage{graphicx}

\shorttitle{Unifying boxy bulge and long bar in the Milky Way}
\shortauthors{Inma~Martinez-Valpuesta \& Ortwin~Gerhard}

\begin{document}

\title{Unifying boxy bulge and planar long bar in the Milky Way}

\author{Inma Martinez-Valpuesta\altaffilmark{1} and Ortwin Gerhard\altaffilmark{1}}
\affil{Max-Planck-Institut f\"ur Extraterrestrische Physik, Giessenbachstrasse, 85748 Garching, Germany}


\altaffiltext{1}{{\it e-mail}: imv  @  mpe.mpg.de, gerhard @ mpe.mpg.de}

\begin{abstract}
  It has been known for some time that the Milky Way is a barred disk
  galaxy. More recently several studies inferred from starcount
  observations that the Galaxy must contain a separate, new, flat {\it
    long bar} component, twisted relative to the barred bulge.  Here
  we use a simulation with a {\it boxy bulge and bar} to suggest that
  these observations can be reproduced with a single structure. In
  this simulation a stellar bar evolved from the disk, and the boxy
  bulge originated from it through secular evolution and the buckling
  instability. We calculate starcount distributions for this model at
  different longitudes and latitudes, in a similar way as observers
  have done for resolved starcounts. Good agreement between the
  simulation and the observations can be achieved for a suitable
  snapshot, even though the simulation has a single {\it boxy bulge
    and bar} structure. In this model, part of the {\it long bar}
  signature is due to a volume effect in the starcounts, and another
  part is due to chosing a snapshot in which the planar part of the
  {\it boxy bulge and bar} has developed leading ends through
  interaction with the adjacent spiral arm heads.  We also provide
  predictions from this model for the line-of-sight velocity
  distributions at the longitudes with the {\it long bar} signature,
  for comparison with upcoming surveys.
\end{abstract}

\section{Introduction}

During the last decades it has become clear that our Galaxy
(hereafter, the MW) is a barred system, as first suggested by
\citet{DeVaucouleurs64}. Now we have good evidence from NIR photometry
\citep{Dwek+95,Binney+97}, star counts
\citep{Stanek+97,LopezCorredoira+05}, gas kinematics
\citep{Englmaier+Gerhard99,Fux99}, microlensing
\citep{Hamadache+06} and dynamical effects near the solar circle \citep{Dehnen00, Minchev+07}. 
Several of these works refer to the boxy bulge
which ends at a galactic radius of $\sim 1.5$~kpc, even though in some
models for the observed data \citep[e.g.,][]{Binney+97} it is clear
that the MW bar extends further in the Galactic plane.

Such a single {\it boxy bulge and bar} structure consisting of a boxy
bulge and a planar bar continuation is a characteristic outcome of the
secular evolution of barred galaxies \citep{Athanassoula05}, where the bar eventually buckles
and forms a boxy-bulge \citep[e.g.,][]{Combes+90,Raha+91}.  After the
buckling event the bar resumes its evolution and continues to grow
slowly through angular momentum exchange between bar, disk and dark
matter halo \citep[e.g.][MV06]{Lynden-Bell+Kalnajs72, Athanassoula03,
  Debattista+Sellwood00, Martinez-Valpuesta+06}.

The MW's {\it boxy bulge and bar} has its long axis in the first
quadrant, at an angle $\alpha\sim15^\circ-30^\circ$ with respect to
the Sun-Galactic center line \citep{Gerhard02}. But more recently star
count observations extending to greater longitudes have led to
surprising results \citep[][hereafter B05,C07,C08,C09]{Benjamin+05,
  Cabrera-Lavers+07b,Cabrera-Lavers+08,Churchwell+09}, confirming an
earlier analysis by \citet{Hammersley+00}: these observations
found indications for a separate {\it long bar} with an orientation of
$\alpha'\sim 43^\circ$, extending from longitude $l\simeq 27^\circ$ to
$l\simeq 10^\circ$ ($4.5$~kpc to $1.5$~kpc galactocentric radius), and
therefore coexisting with the conventional Galactic {\it boxy bulge
  and bar} over a range of radii. This interpretation, if correct,
would dynamically be quite puzzling: two separate rotating bars should
align with each other through dynamical coupling in at most a few
rotation periods.
 
In this letter we show that a separate inferred {\it long bar} does
not necessarily follow from the star count data, and we suggest a
plausible model to explain these data with a single barred structure
whose inner parts represent the boxy bulge. We also 
show some predictions for radial velocity distributions that
could be used to test this model in the near future.

\section{A model for the Milky Way's bar and bulge formed through secular evolution}

\begin{figure}[!t]
\begin{centering}
\includegraphics[scale=0.58,angle=-90]{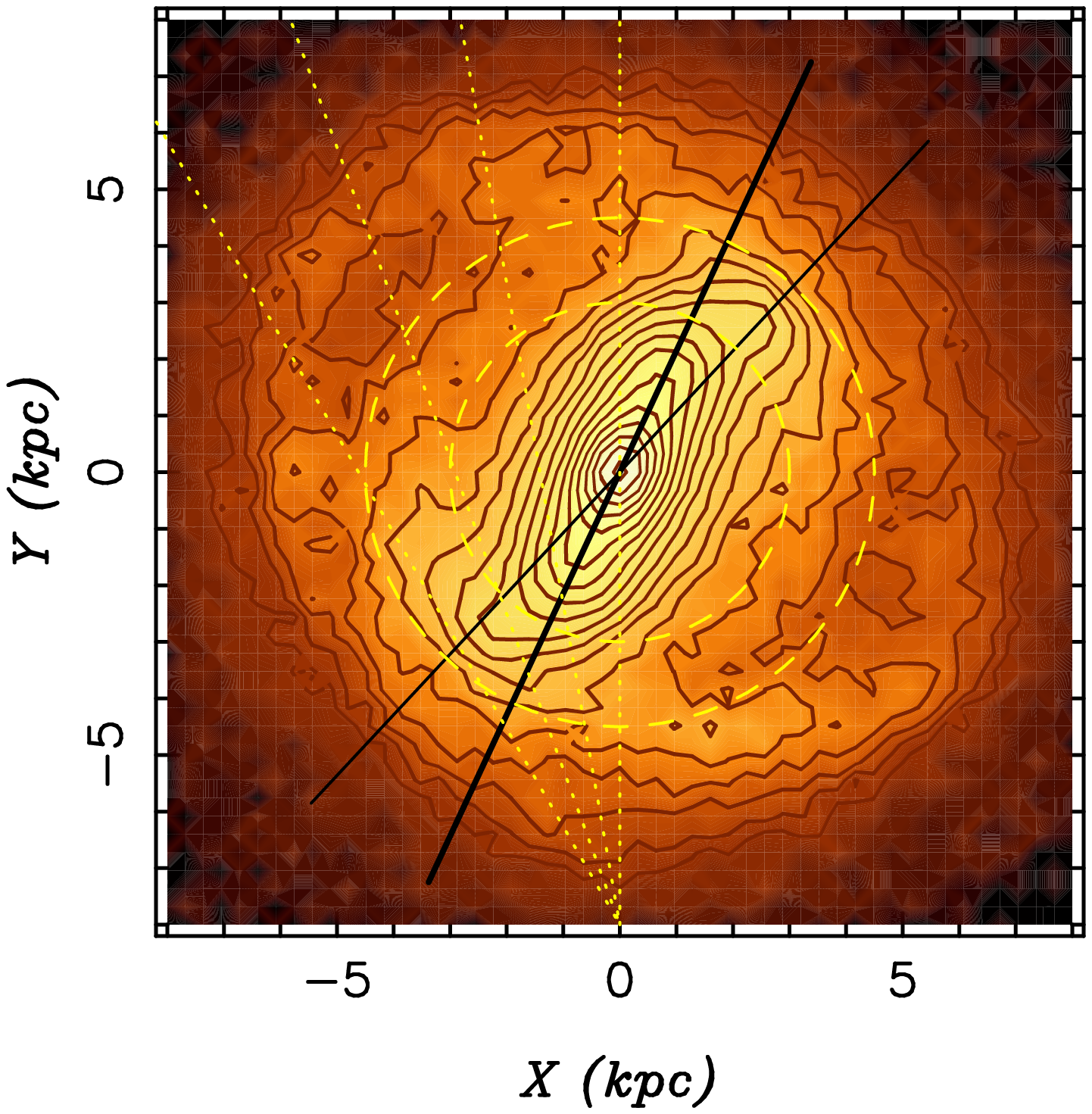}
\includegraphics[scale=0.54,angle=-90]{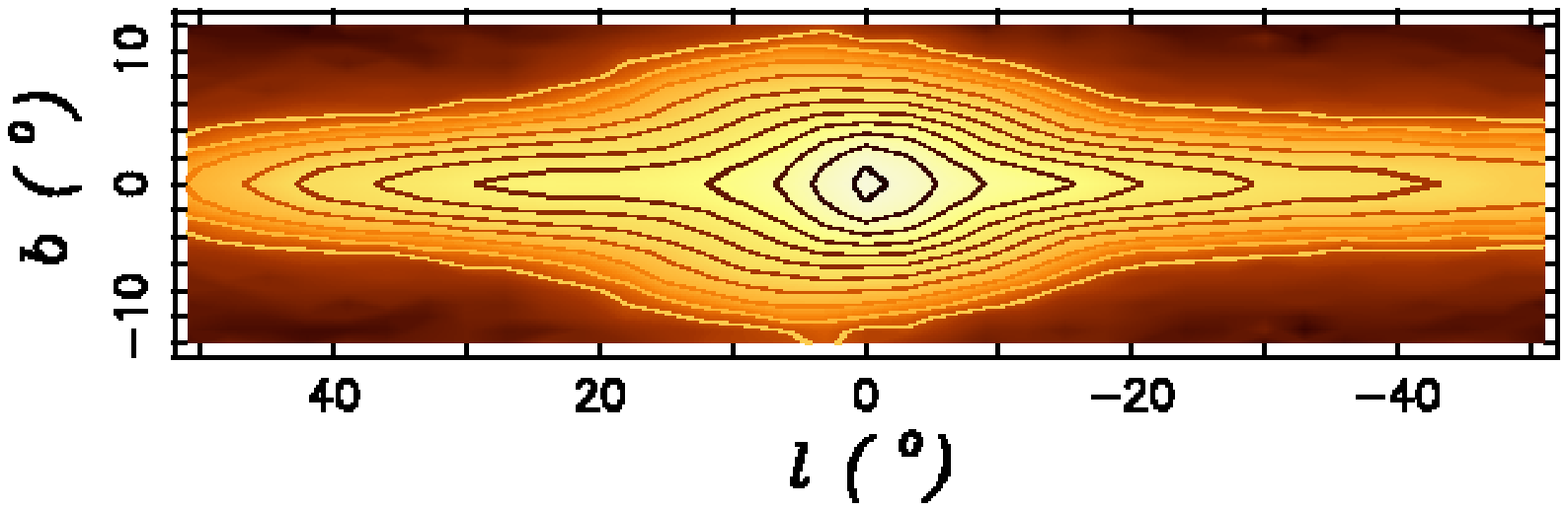}

\end{centering}
\caption{Top panel: Face-on view of the simulation at time $T\sim
  1.9$~Gyr.  The bar rotates clockwise and its ends bend towards the
  leading side, connecting to the spiral arms further out. The model
  has been scaled to the MW and is oriented such that the long axis of
  the bulge is seen at an angle $\alpha=25^\circ$ by the observer at
  (0,-8 kpc). A second line at $43^\circ$ as inferred for the
  {\it long-bar} in the MW is also shown. The two circles correspond to
  radii of $3$ and $4.5$~kpc. The dotted lines show lines-of-sight for
  longitudes of $0^\circ$, $10^\circ$, $20^\circ$ and $30^\circ$ from
  the observer.  Lower panel: edge-on view of the same snapshot, as
  viewed from the Sun. The boxy structure is noticeable. Higher
  densities correspond to brighter colors.}
\label{fig:snapshot1}
\end{figure}

\begin{figure}[!t]
\begin{centering}
\includegraphics[scale=0.58,angle=-90]{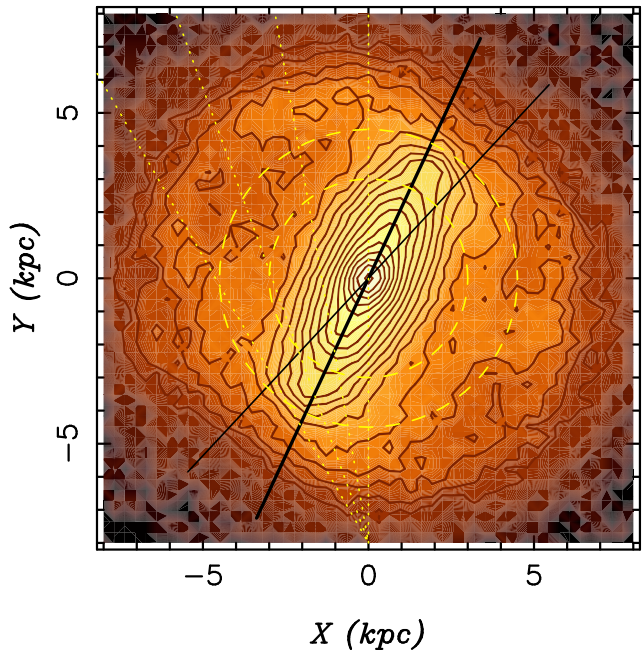}
\end{centering}
\caption{A snapshot from the same simulation, but at this particular time
the ends of the bar are aligned with the bar's inner part. This snapshot is used to illustrate
the volume effect but is not
as good a representation of the MW as that in Figure 1; see Section 2.1.}
\label{fig:snapshot2}
\end{figure}

The simulation used in this work is similar to that published in MV06
and has not been run to match the MW structure. The code used is
FTM~4.4 (updated version) from \citet{Heller+Shlosman94}. The total
number of particles is $1\times10^6$, distributed initially in an
exponential disk with $Q=1.5$, embedded in a live dark matter
halo. After $\sim 1.5$~Gyr the bar becomes very strong and buckles,
thereby weakening. Later the bar resumes its evolution and grows
again, resulting in a prominent {\it boxy bulge and bar} structure.

We consider the simulated galaxy at time $\sim 1.9$~Gyr, after the
boxy bulge has formed and the bar has regrown. The density
distribution for this snapshot is shown in
Figure~\ref{fig:snapshot1}{\it a}, oriented at an angle $\alpha =
25^\circ$ with respect to the line from the Galactic center to the
observer.  The boxy bulge is apparent in
Figure~\ref{fig:snapshot1}{\it b}.  The model is scaled so that the
end of the planar bar appears just inside longitude $l=30^\circ$ as
seen from the observer.  The bar length is $\sim4.5$~kpc, and the
maximum ellipticity is $0.46$.  Relative to the planar bar, the boxy
bulge is $\sim20\%$ larger in $l$ in the scaled model than in the MW,
as measured by comparing the maxima in the asymmetry map with those
from the COBE data \citep{Bissantz+97,Bissantz+Gerhard02}. Also both
the boxy bulge and the disk are vertically more extended in the
model. The Sun is placed at $8$~kpc.

In the face-on view we can easily identify the {\it curved, leading}
ends of the stellar bar. Over a period of 1.2~Gyr, the model shows oscillations from
leading through straight to trailing ends and back. The bar spends 40\% 
of this time in the leading phases. Similar morphology
can be seen in other barred simulations in the literature
\citep[e.g.][model m08]{Fux97} and also in some observed galaxies such
as NGC~3124 \citep{Efremov11} and NGC~3450 \citep{Buta+07}.  The
oscillations between trailing and leading ends of the bar could be
related to the oscillations seen in the bar growth in N-body
simulations \citep[e.g.][]{Dubinski+09} and may be due to non-linear
coupling modes between the bar and spiral arms \citep{Tagger+87}.
This topic is beyond the scope of this paper.  For comparison we show
a snapshot at a later time in this simulation where the ends of the
bar are straight and the spiral arms appear to emerge from them
(Figure~\ref{fig:snapshot2}).

\subsection{Quantitative analysis: no separate 'long bar' is needed}

We apply a similar technique as was used to identify two different
barred structures in the MW from star count data (B05,C07,C08,C09). We
view the projected model as an observer in the disk at $8$~kpc
distance from the center would see it. To increase the particle
resolution, we symmetrize the model vertically and divide the
{\it (longitude, latitude)}-($l,b$)-space into bins of $\delta
l=3^\circ$ and $\delta b=2^\circ$, respectively. Then we count
particles in each of the corresponding cones and bin these particles
in distance modulus, with $\delta\mu = 0.1$. The distance modulus is
given by $\mu=-5.+5.\times\log(D[pc])$.

\begin{figure}[!t]
\begin{centering}
\includegraphics[scale=0.65,angle=-90.]{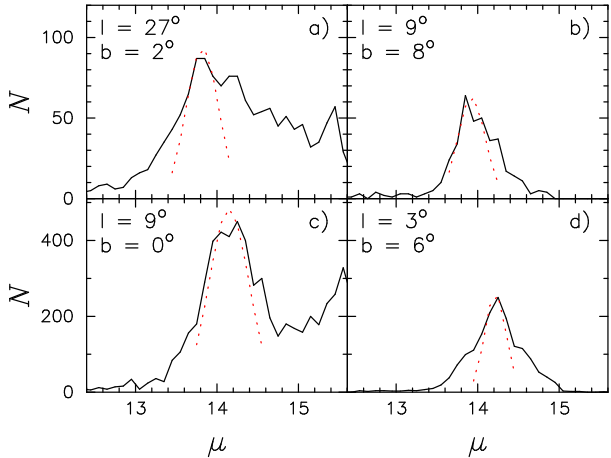}
\end{centering}
\caption{Distribution  of  particles  with  distance modulus  in  four
  fields as seen by an observer at the Sun's position $8$ kpc from the
  center.  The  left panels  {\it  a}),  {\it  c}) show  distributions
  in/near the Galactic plane in  cones centered on the given longitude
  and latitude.  The right panels  {\it b}), {\it d})  show histograms
  for cones through the boxy bulge. }
\label{fig:histograms}
\end{figure}

In Figure~\ref{fig:histograms} we show histograms in $\mu$ for several
lines-of-sight. To quantify the distribution of particles with $\mu$
and to assign a distance value to the maximum number counts, we fit a
Gaussian to the left-most peak, i.e., the one nearest to the
observer.  In the histogram obtained when looking towards the ends of
the bar in the Galactic plane (Figure~\ref{fig:histograms}{\it a}) we
can identify three main peaks, one corresponding to the bar, one to a
spiral arm in the back, and one to the end of the disk. The
$\mu$-value of the fitted maximum corresponds to the end of the bar,
where the bar is flat. The second histogram for
$(l,b)=(9^\circ,8^\circ)$ (Figure~\ref{fig:histograms}{\it b}) shows
the distribution of stars in a field well above the plane where the
boxy bulge dominates. The fitted maximum corresponds to a position on
the thick line in Figure~\ref{fig:snapshot1} at $\alpha=25^\circ$.
In the first panel we can clearly
identify the particles in the disk, but in the second showing the
higher latitude field, disk particles are absent.  We also show the
histogram for $(9^\circ,0^\circ)$ (Figure~\ref{fig:histograms}{\it
  c}), where we can see the increment in the number of particles with
respect to those at $(27^\circ,2^\circ)$, and the displacement of the
maximum towards larger distance in comparison with
$(9^\circ,8^\circ)$. In the last panel we show one of the central
lines-of-sight $(3^\circ,6^\circ)$ for the boxy bulge.

\begin{figure}[!t]
\begin{centering}
\includegraphics[scale=1.2,angle=-90.]{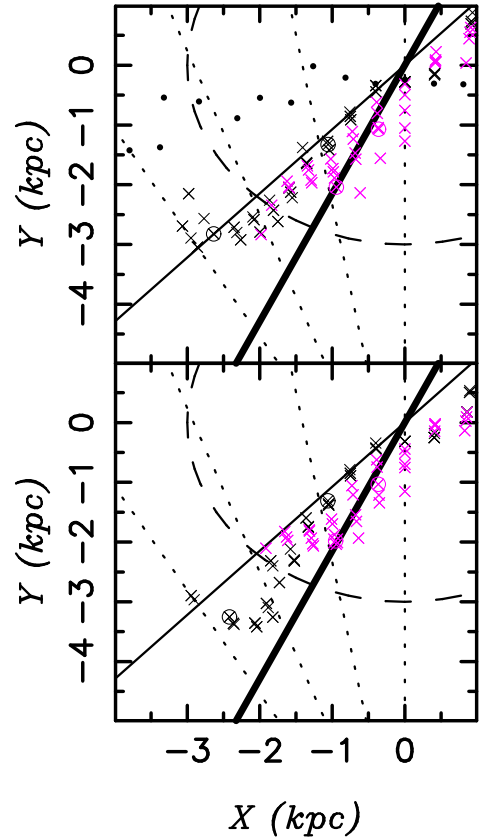}
\caption{
  Location of the star count maxima in the Galactic plane, for fields 
  near the disk plane (black crosses) and in the boxy bulge
  ($4^\circ \leq |b| \leq 8^\circ$, pink crosses). The
  top panel (a) shows the maxima for the model with leading curved
  ends of the bar; black dots show the maxima
  for the initial axisymmetric disk. The lower panel (b) is for the model with straight
  bar ends.  The circled crosses
  correspond to the histograms shown in Figure~\ref{fig:histograms}. 
  The thick solid line shows the true orientation of the
  model, $\alpha=25^\circ$. The thin line follows
  $\alpha'=43^\circ$. The dashed circle has a radius of 3 kpc.}
\label{fig:posdiagram}
\end{centering}
\end{figure}

Repeating the identification of maxima for all fields in
$-6^\circ \le l \le 30^\circ$ and $-8^\circ \le b \le 8^\circ$,
we produce a similar plot as in C07. The distance modulus for each
maximum is converted to distance in kpc from the observer's position
and plotted in Figure~\ref{fig:posdiagram} in galactocentric
coordinates. We show such a plot both for the snapshot with leading
bar ends (Figure~\ref{fig:posdiagram}{\it a}, corresponding to
Figure~\ref{fig:snapshot1}), and for the snapshot with straight ends
(Figure~\ref{fig:posdiagram}{\it b}, corresponding to
Figure~\ref{fig:snapshot2}). In both models, the points computed from
star counts above the plane (pink crosses) follow the heavy line
indicating the true orientation angle $\alpha=25^\circ$ fixed by us
for the model's {\it boxy bulge and bar}, except at the very center
and at negative $l$ where the maxima are closer to the observer due to
tangent point effects.

For lower latitudes (black crosses), the points move further back.  In
both models, for longitudes $0^\circ<l<9^\circ$ in the region of the
boxy bulge, points in the plane now approximately follow an imaginary
line with $\alpha'=43^\circ$. This is due to a volume effect combined
with the shallow disk density distribution: $N(\mu)\propto n(D)D^2 dD/d\mu$.
There is an intermediate
region with $9^\circ<l<20^\circ$, where the boxy bulge becomes thinner
and transits into the planar bar, and then there is a region,
$20^\circ<l<30^\circ$ ($\sim 3$ to $\sim 5$~kpc), where we can see
just particles in the plane. In this region, both models differ: the
volume effect alone, as illustrated for the bar with the straight ends
in Figure~\ref{fig:posdiagram}{\it b}, accounts for only half the shift
from $\alpha=25^\circ$ to $\alpha'=43^\circ$. Whereas for the snapshot
with the leading bar ends (Figure~\ref{fig:posdiagram}{\it a}) the
in-plane points now approximately follow the imaginary
$\alpha'=43^\circ$ line right until the end of the bar is reached.  If
we were not aware of the original structure of the model, the results
shown in Figure~\ref{fig:posdiagram} could easily be interpreted as
evidence for a model with two structures, one thicker and shorter at
 $\alpha = 25^\circ$ and another thinner and longer at $\alpha'\sim 43^\circ$.
Clearly, the case discussed shows that a separate {\it long bar} is not
necessarily implied by the observed star count distributions and that
a single {\it boxy bulge and bar} structure is a valid and dynamically
simpler interpretation of these data.

\section{Radial velocity predictions}

\begin{figure}[!t]
\begin{centering}
\includegraphics[scale=0.50,angle=-90.]{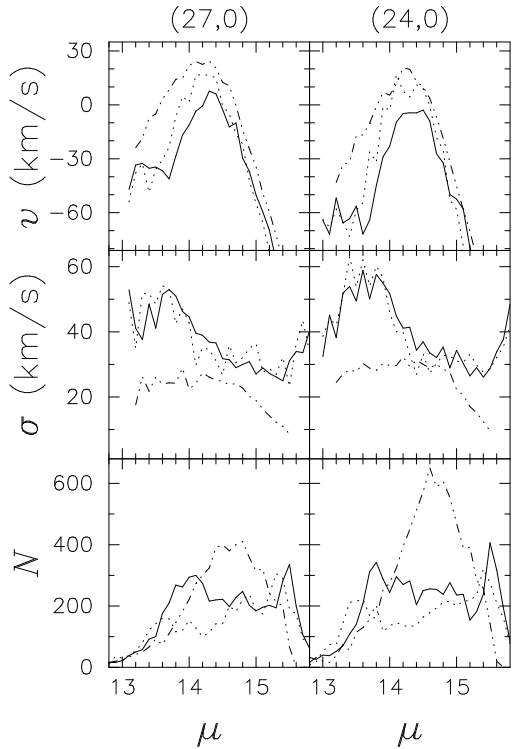}
\end{centering}
\caption{Dependence of mean radial velocity (top), velocity dispersion
  (middle), and particle number (bottom) on distance modulus, for
  two fields centered at $27^\circ$ and at $24^\circ$. Solid, dotted,
  and dot-dashed lines are for the snapshot with leading-ends of the
  bar, for the bar with straight ends, and for the initial exponential
  rotating disk, respectively. The signature of the bar can be easily
  identified; see text.}
\label{fig:velhisto}
\end{figure}

Several large radial velocity surveys of Galactic bulge and disk stars
are currently on-going or planned, and we thus provide some
predictions for our model.  We scale the average circular velocity
curve of the model to the data of \citet{Clemens85} for radii 3-4.5
kpc, and then compute the radial velocities at different longitudes as
seen from the Sun \citep[assuming $v_\sun$=250 km/s;][]{Reid+09}. We
show the mean velocity, dispersion, and particle number vs.\ distance
modulus in Figure~\ref{fig:velhisto} for two in-plane positions near
the end of the planar bar. The non-circular motions in the barred
models leave a clear signature at $\mu\simeq 13.0-13.6$ compared with
the initial axisymmetric disk. In this range the
bar's velocity profile is flat. For the leading-ends bar
it has a break at $\mu\simeq13.7$ and then increases towards an
axisymmetric type of profile. For the straight-ends bar
the break happens earlier $\mu\simeq13.5$ and the radial velocities
are slightly smaller. The difference between axisymmetric and bar case
is $\simeq 30-40$~km/s. The velocity dispersions of both bar models
in the two fields are higher by $\sim 20$~km/s for all $\mu<14.2$.
Also note the difference between the star counts with $\mu$ between the three
models.

Velocities in the boxy bulge are outside the scope of this paper; they
have already been studied with a similar N-body simulation by
\citet{Shen+10} and compared with results from the BRAVA survey.

\section{Summary and Conclusions}

We have analyzed star counts in the inner Galaxy using an N-body model
which arose from secular evolution of a disk galaxy, a natural
mechanism for galaxies like ours. The bar and buckling instabilities
in the stellar disk lead to a boxy bulge which extends to a longer
in-plane bar. The bar couples with the spiral arms in the disk, giving
rise alternately to leading, straight or trailing bar ends.

As seen from within the disk at 8 kpc from the center, the maxima of
the line-of-sight distance distributions in the Galactic plane occur
at distances somewhat further than the maxima of the line-of-sight
density distributions, due to the volume effect in the star
counts. Assuming a plausible orientation ($\alpha=25^\circ$), this
explains part of the observational signature which was previously used
to infer the existence of a second {\it long bar}.  If in addition we
choose a model snapshot where the bar has leading ends, most of the
{\it long bar} signature in the star count data can be reproduced.
While not made specially to match the MW, this model thus illustrates
that the traditional Galactic bar (the boxy bulge) and the more
recently inferred {\it long bar} can plausibly be explained by a
single {\it boxy bulge and bar} structure.

To test this further we have determined the dependence of the mean radial
velocity and velocity dispersion on distance modulus in Galactic plane fields near the
inferred end of the planar bar. These illustrate the differences
between the barred model and an axisymmetric rotating disk, which can
be compared with upcoming radial velocity survey data for the inner
MW. Future work should also address in more detail the spiral arm -
bar interaction which gives rise to the curved ends of the bar, and
aim at constructing a more detailed dynamical model allowing us to
understand better the structure and evolution of the inner Milky Way.
 
\section*{Acknowledgments}
We thank Ken Freeman for helpful discussions about this work. IMV
thanks Peter Hammersley for discussions on the MW.

\bibliographystyle{apj}

\end{document}